# Singlet fission and tandem solar cells reduce thermal degradation and enhance lifespan

Y. Jiang[1], M. P. Nielsen[1], A. J. Baldacchino[1], M. A. Green[1], D. R. McCamey[2], M. J. Y. Tayebjee[1], T. W. Schmidt[3] and N. J. Ekins-Daukes[1]

[1]School of Photovoltaic and Renewable Energy Engineering, UNSW Sydney, NSW 2052, Australia
[2]ARC Centre of Excellence in Exciton Science, School of Physics, UNSW Sydney, NSW 2052, Australia
[3]ARC Centre of Excellence in Exciton Science, School of Chemistry, UNSW Sydney, NSW 2052, Australia

## Abstract

The economic value of a photovoltaic installation depends upon both its lifetime and power conversion efficiency. Progress towards the latter includes mechanisms to circumvent the Shockley-Queisser limit, such as tandem designs and multiple exciton generation (MEG). Here we explain how both silicon tandem and MEG enhanced silicon cell architectures result in lower cell operating temperatures, increasing the device lifetime compared to standard c-Si cells. Also demonstrated are further advantages from MEG enhanced silicon cells: (i) the device architecture can completely circumvent the need for current-matching; and (ii) upon degradation, tetracene, a candidate singlet fission (a form of MEG) material, is transparent to the solar spectrum. The combination of (i) and (ii) mean that the primary silicon device will continue to operate with reasonable efficiency even if the singlet fission layer degrades. The lifespan advantages of singlet fission enhanced silicon cells, from a module perspective, are compared favorably alongside the highly regarded perovskite/silicon tandem and conventional c-Si modules.

At present, wafer-based silicon modules are the dominant PV technology. These modules are currently provided with warrantied lifetimes usually of 25 years. Progress in all areas of PV module manufacturing will reduce costs, but there remains significant scope to reduce the cost of PV electricity through improved power conversion efficiency and by increasing the module lifespan beyond the present 25 years.

The power conversion efficiency of conventional cells are approaching the single threshold Shockley-Queisser (SQ) efficiency limit[1,2]. Efforts to circumvent this limit include the introduction of multiple absorbing thresholds, which can take several forms. Tandem cells incorporate a second junction with a larger bandgap material (such as perovskites or III-Vs), and excitonic methods exploit multiple-exciton generation processes, such as singlet fission. The process of singlet fission in molecular semiconductors is well established, whereby a photo-excited singlet exciton undergoes fission into two triplet excitons, producing twice the electronic charge carriers for each absorbed photon.

While both these approaches increase efficiency, little thought has been given to their impact on thermal load, particularly when considering the degradation present in realistic devices. Here, we provide a model for understanding the impact of multiple absorbing thresholds on thermal load. We investigate the thermal load reduction from both tandem and singlet fission cells, and estimate the resulting improvement to the lifespan of a module made from those cells.

**Solar module heat flow analyses**

A schematic of the heat generation and loss processes in a solar module is shown in Fig. 1. A portion of the power of the incident sunlight is absorbed in the module, largely in the cells rather than the encapsulant. The encapsulation is an optically transparent glass coversheet with a thin absorbing rear polymeric coversheet placed against the cells. Taking into account optical reflection, the power into the module ($P_{in}$) is reduced compared to the incident sunlight. This heat is conducted to both module surfaces from where it is dissipated by free ($P_{c,free}$) and forced ($P_{c,forced}$) convection as well as radiatively ($P_{rad.}$). However, it should be noted that encapsulating glass is largely opaque at the radiative wavelengths of silicon[3].

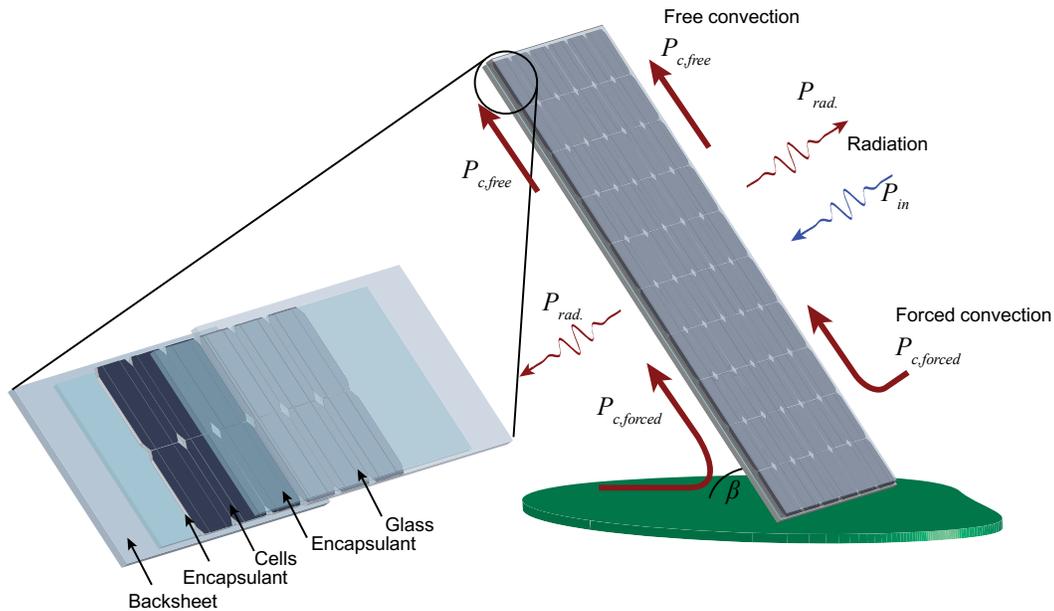

**Fig. 1 | Solar module energy balance.** Incoming radiation $P_{in}$ results in electrical power $P_{elec.}$ and heat losses via radiative emission ($P_{rad.}$), forced ($P_{c,forced}$) and free ($P_{c,free}$) convection.

Prior to 2016, module thermal performance was characterised in terms of a single parameter, the nominal operating cell temperature (NOCT), defined as the temperature reached by open circuited cells in a module under a defined set of operating conditions: irradiance of 800 W·m$^{-2}$, 20°C ambient temperature, wind speed of 1 m·s$^{-1}$, and mount tilt angle of 45°. Following the introduction of an updated standard (IEC61215: 2016), NOCT has now been replaced by the NMOT (nominal module operating temperature), which is defined slightly differently[4]. Consistent with the name change, the back-of-module temperature is now measured rather than the cell temperature. The standard reference environment remains unchanged with one major caveat, the electrical operating conditions are set to when the module load matches peak power generation instead of open-circuit. Experimental differences are small, with measurements taken at NREL (National Renewable Energy Laboratory, USA) over 38 days for a single module[4] giving NMOT = 48.7±1.7°C and NOCT = 47.9±1.3°C.

In our thermal model for c-Si based devices (see Methods section for details), solar photons absorbed above the bandgap ($\lambda$ < 1200nm) convert their energy to both electricity and excess heat generation, while all photons absorbed below the bandgap contribute to heat only[5]. Applying these assumptions to our thermal model, the thermalization losses within a solar module become $P_{in} - P_{elec.}$. The solar module operating temperature was determined by solving $P_{in} - P_{elec.} = P_{rad.} + P_{convec}$ where these parameters are defined in Fig. 1. To validate the model, the working voltage for a hypothetical solar cell is estimated with $V_{mp\_Si}$ = 0.6V for a c-Si solar cell. Solving the energy balance equation results in $T_{module}$ = 42.4°C, in line with the rule of thumb that modules operate 20-30°C above the ambient temperature.

We next apply the model to two well-studied approaches to multiple threshold devices: a perovskite on silicon tandem and a tetracene-based singlet fission device. For the perovskite/Si tandem cell, the working voltage is taken as 1.4V (0.8V and 0.6V for the perovskite and Si junctions, respectively)[6]. Singlet fission in the molecular semiconductor tetracene is well-known to generate two triplet excitons that are energetically matched to the silicon bandgap. The singlet fission energy threshold is assumed to be at 530nm, slightly endothermic with $\Delta E$=0.2eV [7]. The thermal load spectra for the different cell architectures is shown in Fig. 2.

The operating temperature of these two configurations are calculated as $T_{PSK/Si}$ = 40.9°C and $T_{SF/Si}$ = 40.4°C, showing a reduction in temperature $\Delta T$= 1.5°C and $\Delta T$ = 2°C compared with the conventional Si module, respectively. c-Si module lifetime is generally found to double for every 10°C reduction in temperature[8], for a thermally activated process this corresponds to an activation energy $E_a = 0.61$ eV at a module temperature of 42.4°C. The lifespan of a module can therefore be expressed using an Arrhenius equation lifespan $= A \times e^{\frac{E_a}{kT}}$, with $A = 4.0 \times 10^{-9}$ years. This equates to an increase in lifetime of 2.7 years (11%) for the perovskite tandem and 3.7 years (14.9%) for the tetracene-based SF cell. A comparison of key parameters for these three structures discussed above are shown in Table 1.

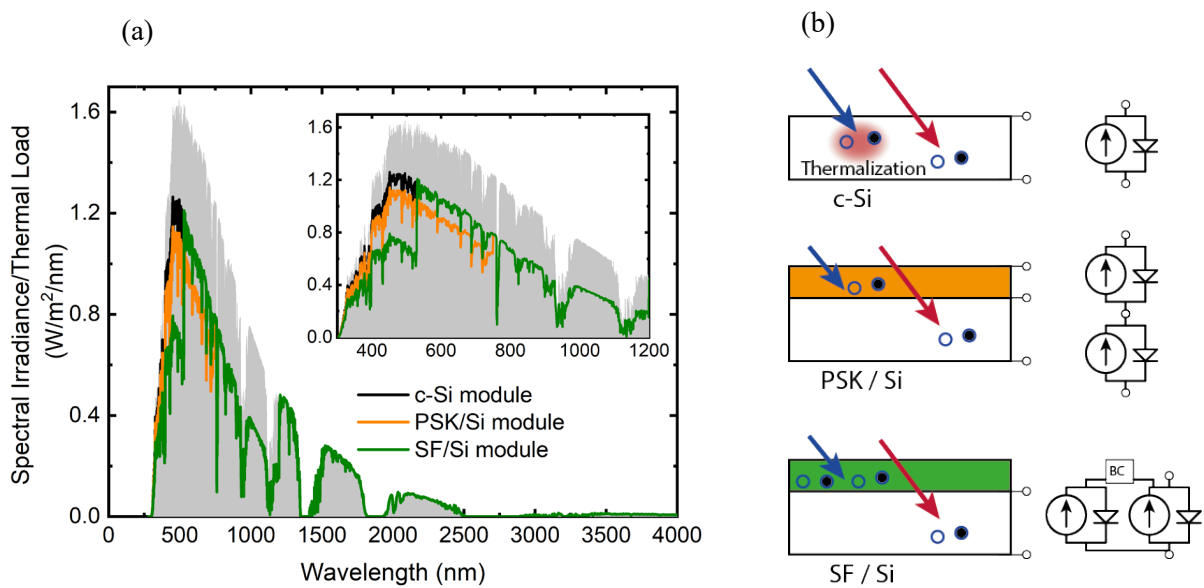

**Fig. 2 | Comparing conventional and tandem devices.** (a) Solar spectral irradiance AM1.5G overlaid with the heat load spectra for various cell architectures. The inset shows the relevant 300-1200nm wavelength range. (b) The carrier generation process in each configuration and the equivalent circuit. BC denotes a buck converter which is a power converter that halves the voltage and doubles the current, representative of the singlet fission process.

**Table 1** | The energy distribution of a solar device for conventional c-Si module, monolithic perovskite/Si module, and singlet fission/Si module.

| Structure | Initial Efficiency (%) | $P_{elelctr.}$ (W) | $P_{rad.}$ (W) | $P_{conv.}$ (W) | Temperature (°C) | Lifespan (years) |
|---|---|---|---|---|---|---|
| c-Si module | 26.5 | 211.7 | 253.5 | 295.1 | 42.4 | 25 |
| PSK/Si module | 31.0 | 248.1 | 237.9 | 274.4 | 40.9 | 27.7 |
| SF/Si module | 32.8 | 262.3 | 231.7 | 266.3 | 40.4 | 28.7 |

## Degradation

Degradation is inevitable in all components of a PV module. Conventional silicon modules operating in the field typically suffer a 0.4% per annum degradation rate; commercial thin-film technologies (CdTe, CIGS) typically degrade roughly twice as quickly[9].

While it is difficult to realistically predict actual degradation rates for new top materials at this stage, perovskite materials are known to suffer from stability issues that require, at a minimum, very effective encapsulation; the same will also be true for tetracene-derived singlet-fission materials. A recent study on the efficiency premium offered by a perovskite silicon tandem established a 3.5% per annum degradation rate as the maximum acceptable value to achieve an overall energy yield advantage[10] compared with conventional silicon cells.

Degradation of singlet-fission layers is recognised as one of the primary obstacles to the development of a commercially viable device. Although the current matching considerations and the consequences of top cell degradation seen in two-terminal tandem series multi-junction devices are not present in the singlet fission device, intrinsic photodegradation can still play a significant role in the commercial viability. It should be noted that the perovskite/Si tandem cell considered here, as a three-terminal device, is not subject to current matching considerations. A two-terminal perovskite/Si tandem cell would have significantly more complicated degradation-related temperature properties as degradation of the top layer would reduce the overall current flow, leading to increased thermalization in the c-Si.

Whilst considerable efforts are underway to identify stable alternatives[11–14], tetracene is currently the only singlet fission material to have demonstrated energy transfer into silicon[15]. Importantly, ditetracene (the degradation product of tetracene under anaerobic conditions[16]) is transparent to solar radiation (Fig. 3). Therefore, although degradation occurs, a singlet fission/silicon solar cell will return to the primary underlying silicon solar cell, having gained both efficiency and lifetime benefits for the stable duration of the tetracene layer. This compares favourably against two-terminal tandem series multi-junction devices where current matching considerations ensure that degradation of the top cell significantly impacts device performance.

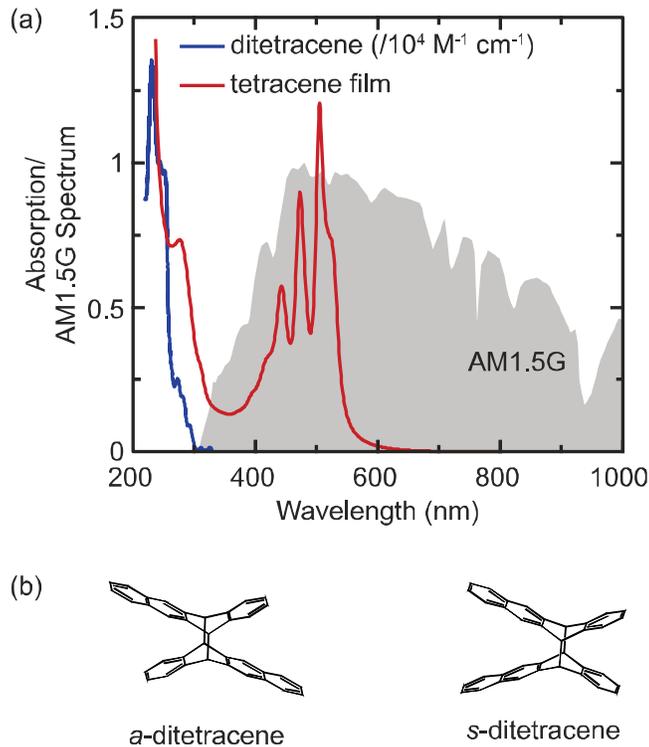

**Fig.3 | Degradation of tetracene leaves a fully operational primary device.** (a) Absorption spectra of isolated tetracene degradation product (from Ref. [17]) and a tetracene thin film (arb. units). The normalised AM1.5G spectrum is shown for reference. (b) The structure of the asymmetric (*a*-) and symmetric (*s*-) ditetracene isomers.

In Fig. 4(a) we show the effect of degradation on module temperature for conventional silicon, perovskite/silicon, and singlet fission/silicon solar modules with a lower bound for degradation at 0.5%/year and an upper bound of 3.5%/year[10] for the perovskite and singlet fission components. At the beginning of life, all cells would sit at the lowest temperature points, corresponding to the lower left-hand side of Fig. 4(a). As time progresses, these cells will degrade at variable rates, but not exceeding the bounds of the shaded region to some end of life state. Since the degradation product of tetracene singlet fission film is transparent to solar radiation, it will revert to standard c-Si solar cells after the benign degradation process, and the temperature of tetracene singlet fission/Si module converges to that of the conventional c-Si module.

Furthermore, the cumulative degradation effect over the lifespan of a solar module on its operating temperature for perovskite/silicon tandem, generic singlet fission/silicon (where the degradation product is not transparent to solar radiation), and tetracene singlet fission/silicon (where the degradation product is transparent to solar radiation) modules are also modelled and shown in Fig. 4(b)-(d) respectively. c-Si solar modules typically suffer < 20% cumulative degradation during a 25 year lifespan, while up to 100% degradation is modelled for perovskite and singlet fission materials in each configuration.

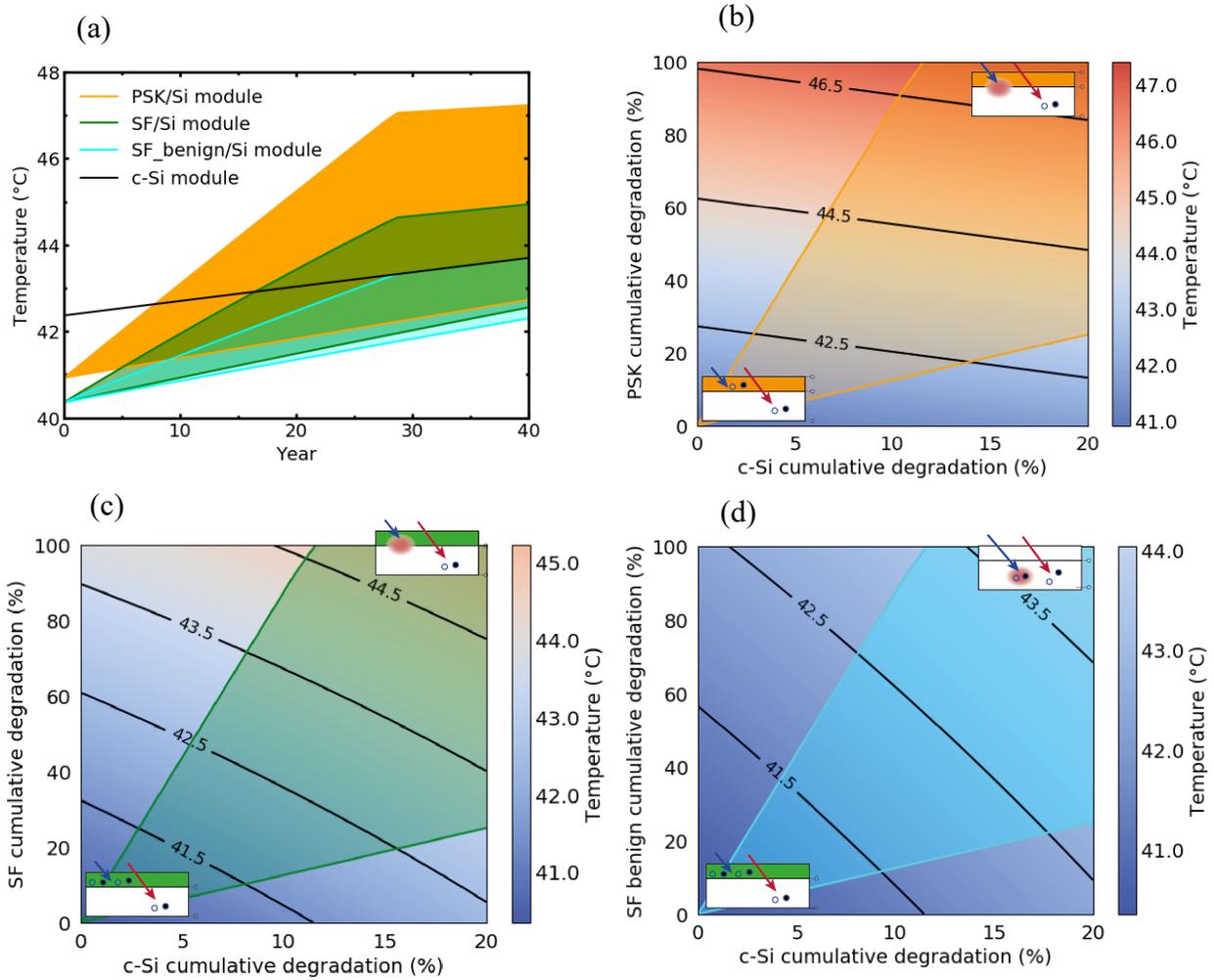

**Fig. 4 | The operating temperature of different device architectures.** (a) Increase in module temperature plotted for modules composed of conventional c-Si, singlet fission/silicon, and perovskite/silicon modules. The c-Si degradation rate is 0.4% per annum; singlet fission/silicon and perovskite/silicon degradation rates have a lower bound of 0.5% per annum and an upper bound of 3.5% per annum. (b)-(d) module operating temperature resulting from cumulative degradation. The inset diagrams depict the carrier generation processes for beginning and end of life for (b) perovskite/silicon, (c) general singlet fission/silicon and (d) tetracene singlet fission/silicon modules respectively. The shaded region corresponds to the degradation rate of 0.5%-3.5% per annum.

# Conclusions

The opportunity for multiple threshold silicon solar cells to increase the spectral efficiency compared with conventional silicon solar cells is well known. Here, we have shown that there are also ancillary benefits to the approach in terms of lower module temperature and resilience under degradation. At low degradation rates, both perovskite-based tandems and singlet-fission cells reduce thermal load and increase the lifetime of the primary silicon module. At larger degradation rates, tetracene-based singlet fission cells still outperform conventional cells as the benign nature of the degradation product is not detrimental to the performance of the primary silicon cell. In all these cases there exists the potential to significantly reduce the cost of energy produced by solar photovoltaic systems by both increasing efficiency and lifespan.

# Methods

**Solar module heat flow analyses.**

The total incident irradiated solar power absorbed by a photovoltaic cell is given by:

$$P_{in} = \int_0^\infty \alpha(\lambda) \times b_{AM1.5G}(\lambda)d\lambda \tag{1}$$

where $b_{AM1.5G}$ is the solar spectral irradiance for sunlight under the standard air-mass 1.5 global condition[18], thereby giving a total irradiance of 1000 W·m$^{-2}$. For a c-Si solar photovoltaic module, the absorptivity $\alpha(\lambda)$ is well approximated by 95% across the entire solar spectrum[14], and accounting for the defined incident irradiance for NMOT (800 W·m$^{-2}$), so the total irradiated solar power becomes:

$$P_{in} = 0.95 \times 0.8 \int_0^\infty b_{AM1.5G}(\lambda)d\lambda \tag{2}$$

The outgoing power comprises contributions from electrical power generation, grey-body radiation, and convection, resulting in:

$$P_{out} = P_{elec.} + P_{rad.} + P_{convec.} \tag{3}$$

The electrical power delivered by the Si cell $P_{elec.Si}$ is given by:

$$P_{elec.Si} = \int_0^{\lambda(E_g)} (1 - d_{Si}) N_{in}(\lambda) V_{mp_{Si}} \, d\lambda \tag{4}$$

where $V_{mp\_Si}$ is the voltage at the maximum power, estimated as 0.6V. $(1 - d_{Si}) N_{in}(\lambda)$ represents the photo generated current, $d_{Si}$ accounts for power degradation of Si module, $N_{in}(\lambda)$ is the spectral photon flux, defined below:

$$N_{in}(\lambda) = 0.95 \times 0.8 \frac{b_{AM1.5G}(\lambda)}{hc/\lambda} \tag{5}$$

h is the Planck constant, and c is the speed of light.

The generation of two electron-hole pairs in the singlet fission/Si cell with benign degradation is accounted with the additional $(1 - d_{Si})(1 - d_{SF}) N_{in}(\lambda)$ term, with degradation of the singlet fission accounted with $d_{SF}$. Endothermic singlet fission process accounted by $(1 - d_{SF}) N_{in}(\lambda) \Delta E$. At energies below the singlet fission threshold, photon generation proceeds as for the conventional Si solar cell.

$$P_{elec.SF_{benign}/Si} = \int_0^{\lambda(E_{g,SF})} (1 - d_{Si}) N_{in}(\lambda) V_{mp_{Si}} + (1 - d_{SF}) N_{in}(\lambda) \Delta E + (1 - d_{Si})(1 - d_{SF}) N_{in}(\lambda) V_{mp_{Si}} \, d\lambda + \int_{\lambda(E_{g,SF})}^{\lambda(E_g)} (1 - d_{Si}) N_{in}(\lambda) V_{mp_{Si}} \, d\lambda \tag{6}$$

In the case of general singlet fission layer degradation, all the photon generation above the singlet fission threshold is subjected to the singlet fission degradation rate.

$$P_{elec.SF/Si} = \int_0^{\lambda(E_{g,SF})} 2 \times (1 - d_{Si})(1 - d_{SF}) N_{in}(\lambda) V_{mp_{Si}} + (1 - d_{SF}) N_{in}(\lambda) \Delta E \, d\lambda + \int_{\lambda(E_{g,SF})}^{\lambda(E_g)} (1 - d_{Si}) N_{in}(\lambda) V_{mp_{Si}} \, d\lambda \tag{7}$$

For the perovskite/Si tandem cell, independent degradation rates are used, $d_{PSK}$ describing the degradation rate of perovskite layer. $V_{mp\_PSK}$ is the perovskite maximum power voltage 0.8V.

$$P_{elec.PSK/Si} = \int_0^{\lambda(E_{g,PSK})} (1 - d_{PSK}) N_{in}(\lambda) V_{mp_{PSK}} \, d\lambda + \int_{\lambda(E_{g,PSK})}^{\lambda(E_g)} (1 - d_{Si}) N_{in}(\lambda) V_{mp_{Si}} \, d\lambda \tag{8}$$

The net radiative power loss will depend on the orientation of the panel with respect to the sky and ground, and their respective temperatures. The net radiative loss per unit module area is given by[19]:

$$P_{rad.} = \frac{1}{2}\sigma(\epsilon_{module,f}\epsilon_{sky}(T_{sky}^4 - T_{module}^4)(1 + \sin(90° - \beta)) + \epsilon_{module,f}\epsilon_{ground}(T_{ground}^4 - T_{module}^4)(1 - \sin(90° - \beta)) + \epsilon_{module,b}\epsilon_{sky}(T_{sky}^4 - T_{module}^4)(1 - \cos\beta) + \epsilon_{module,b}\epsilon_{ground}(T_{ground}^4 - T_{module}^4)(1 + \cos\beta)) \quad (9)$$

where σ is the Stefan–Boltzmann constant, β is the angle of elevation for the module from the horizontal and taken to be 45°, $\epsilon_{module,f}$, $\epsilon_{module,b}$, $\epsilon_{sky}$, and $\epsilon_{ground}$ are emissivity values for front of module, back of module, sky, and ground, taken to be 0.84, 0.893[4], 0.82[20], and 0.95[21], respectively, and the sky temperature $T_{sky}$ is calculated by[22]:

$$T_{sky} = 0.0552 \times T_{ambient}^{1.5} \quad (10)$$

Convective heat transfer per unit module area is expressed by:

$$P_{convec.} = P_{c,forced} + P_{c,free} = -(h_{c,forced} + h_{c,free})(T_{module} - T_{ambient}) \quad (11)$$

where the heat transfer coefficient for forced convection at wind speeds of 1 m·s⁻¹ is assumed to be 9.5 W·m⁻²·K⁻¹. The heat transfer coefficient for free convection from a vertical plane in air can be approximated by[23]:

$$h_{c,free} = 1.31 \sqrt[3]{T_{module} - T_{ambient}} \quad (12)$$

## Data availability

The authors declare that the data supporting the findings of this study are available within the paper, Supplementary Information, and Source Data files. Further data beyond the immediate results presented here are available from the corresponding authors upon reasonable request.

## Acknowledgements
Y.J. is supported by an ACAP fellowship. M.P.N is supported by a UNSW Scientia Fellowship. This work was supported by the Australian Research Council (Centre of Excellence in Exciton Science CE170100026).

## Author information

**Contributions**
The ideas were conceived by T.W.S., N.J.E-D and M.A.G. Singlet fission degradation data was supplied by M.J.Y.T., D.R.M. and A.J.B, perovskite data by Y.J. The calculations were performed by Y.J. and verified by T.W.S., N.J.E-D and M.A.G. The first draft of the paper was written by Y.J. with subsequent revision by M.J.Y.T, D.R.M and M.P.N. The figures were prepared by Y.J, N.J.E-D and M.J.Y.T. All authors discussed the results and commented on the manuscript.

C**orresponding authors**
Correspondence to N.Ekins-Daukes nekins@unsw.edu.au

## Ethics declarations

**Competing interests**
The authors declare no competing interests.